\documentclass{article}

\usepackage{arxiv}

\usepackage{balance}
\usepackage[utf8]{inputenc} 
\usepackage[T1]{fontenc}    
\usepackage{hyperref}       
\usepackage{url}            
\usepackage{booktabs}       
\usepackage{amsfonts}       
\usepackage{nicefrac}       
\usepackage{microtype}      
\usepackage{lipsum}
\usepackage{fancyhdr}       
\usepackage{graphicx}       
\usepackage{amsmath}
\graphicspath{{media/}}     

\title{
  {\SBTRec} - A Transformer Framework  for 
  Personalized \\ Tour  Recommendation Problem  
  with Sentiment Analysis} 

\author{
    Ngai Lam Ho,
    Roy Ka-Wei Lee and
    Kwan Hui Lim \\
  Singapore University of Technology and Design \\
  Singapore \\
  \\
  \texttt{1004986@mymail.sutd.edu.sg,
   \{roy\_lee, kwanhui\_lim\}@sutd.edu.sg\} } \\
}

\newcommand{\BTRec}{{\textsc{\text{BtRec}}}}
\newcommand{\SBTRec}{{\textsc{\text{SBtRec}}}}
\newcommand{\POIBert}{{\textsc{{PoiBert}}}}
\newcommand{\PerPoiBert}{{\textsc{\text{PPoiBert}}}}
\newcommand{\BERT}{{\textsc{\text{Bert}}}}
\newcommand{\CBOW}{{\textsc{\text{Cbow}}}}
\newcommand{\SkipGram}{{\textsc{\text{SkipGram}}}}
\newcommand{\SB}{{\textsc{\text{SBert}}}}
\newcommand{\SBert}{{\textsc{\text{SBert}}}}

\newcommand{\spmf}{{\textsc{\text{Spmf}}}}
\newcommand{\SuBSeq}{{\textsc{\text{SuBSeq}}}}
\newcommand{\CPT}{{\textsc{\text{Cpt}}}}
\newcommand{\CPTplus}{{\textsc{\text{Cpt$+$}}}}
\newcommand{\TDAG}{{\textsc{\text{Tdag}}}}

\newcommand{\LSTM}{{\textsc{\text{Lstm}}}}
\newcommand{\LBSN}{{\textsc{LBSN}}}
\newcommand{\POI}{{\textsc{Poi}}}
\newcommand{\POIs}{{\textsc{Poi}s}}
\newcommand{\NLP}{{\textsc{NLP}}}
\newcommand{\MLM}{{\textsc{Mlm}}}
\newcommand{\NSP}{{\textsc{Nsp}}}
\newcommand{\CLS}{{\texttt{[CLS]}}}
\newcommand{\MASK}{{\texttt{[MASK]}}}
\newcommand{\SEP}{{\texttt{[SEP]}}}
\newcommand{\uprofile}{{\bar{u}}}

\newcommand{\F}{{${\mathcal{F}_1}$}}
\newcommand{\Unmask}{\textbf{\textsf{Unmask}}}

\newcommand{\LET}{\textbf{let}}

\newcommand{\WordVec}{{\textsc{\text{WordVec}}}}

\newcommand{\Buda}{{Budapest}}
\newcommand{\Delh}{{Delhi}}
\newcommand{\Edin}{{Edinburgh}}
\newcommand{\Glas}{{Glasgow}}

\newcommand{\Osak}{{Osaka}}
\newcommand{\Pert}{{Perth}}
\newcommand{\Toro}{{Toronto}}
\newcommand{\Vien}{{Vienna}}

\newcommand{\NextPop}{{\textsc{NextPop}}}
\newcommand{\recall}{{${\mathcal{R}}$}}
\newcommand{\precision}{{${\mathcal{P}}$}}

\newcommand{\triplet}[3]{\shortstack{{#1}\\{#2}\\{#3}}}

\newcommand{\notationTab}{
    {\begin{tabular}{cl}
        \hline
        \hline
          ~ & Description \\
        \hline
          $\hat{F_i}$ & \emph{Expected} number of photos at {\POI}-$p_i$\\
        \hline
          $H_u$   & Registered city/country of $u$ \\
        \hline
          $p^i_j$ & $\POI$ in Step-$j$ of $i$'s itinerary \\
        \hline
         $p_u$    & source {\POI} of user's itinerary\\
        \hline
         $p_v$    & destination {\POI}~of user's itinerary\\
        \hline
          $S_h$   & sequence of {\POI}~as a user's itinerary \\
        \hline
          $S_p$   & Predicted {\POI}~itinerary from recommendation \\
        \hline
          $\SB_i$ & Sentence~\BERT~embedding from comments posted to \POI-$i$ \\
        \hline
          $C^i_j$ & Category~label of \POI-$p_i$ \\
          ~ & in step-$j$ of trajectory,  \\
          ~ & e.g. `Sport', `Shopping',.. etc.\\
        \hline
          $T$     & Total time budget allocated \\
        \hline\hline
    \end{tabular}}
}
\newcommand{\CorpusGeneration}{
    {
        \begin{figure}
            \label{alg:CorpusGeneration}
            \includegraphics[width=1.02\textwidth,
              trim=2.3cm 19.1cm 5.6cm 2.3cm,
              clip
              ]{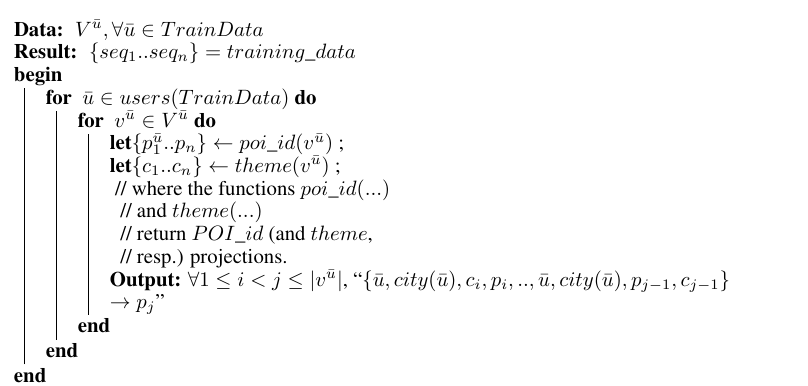}
            \caption{ Generation of Training Data~with users' preference model }
        \end{figure}
    }
}
\newcommand{\SBTRecAlgoOneColumn}{
    {
        \begin{figure*}

        {\includegraphics[width=1.02\textwidth,
                  trim=1.50cm 19.2cm 5.6cm 2.5cm,
                  clip
                  ]{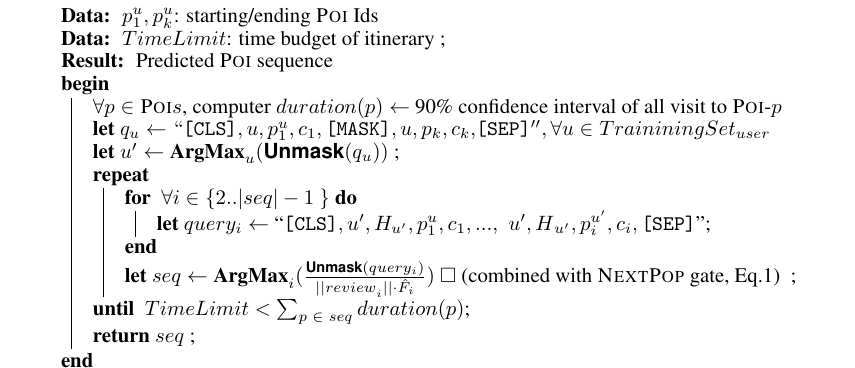}
                  
            \label{alg:SBTRec}
            \caption{Itinerary Prediction Algorithm in {\SBTRec}}
        }
        \end{figure*}
        }
}

\newcommand{\ExptTab}{ 
        {\scalebox{0.95}{
            \begin{tabular}{lccccccccccc}
            \hline\hline
              {\hspace{-1mm}}\textsf{Alg. }{\hspace{-4mm}}&&
              \textsf{\Buda}{\hspace{-4mm}}&
              \textsf{\Delh}{\hspace{-4mm}}&
              \textsf{\Edin}{\hspace{-4mm}}&
              \textsf{\Glas}{\hspace{-4mm}}&
              \textsf{\Osak}{\hspace{-4mm}}&
              \textsf{\Pert}{\hspace{-4mm}}&
              \textsf{\Toro}{\hspace{-4mm}}&
              \textsf{\Vien}{\hspace{-4mm}}&
              {\hspace{-2mm}}\textsl{All cities}{\hspace{-10mm}}\\
            \hline
            \vspace{1mm}\triplet{~}{CPT}{~}&
              \triplet{\recall}{\F}{\precision}&
              \triplet{64.36}{49.69}{63.28} &
              \triplet{82.22}{53.57}{64.45} &
              \triplet{68.38}{51.47}{61.97} &
              \triplet{71.82}{63.88}{71.97} &
              \triplet{58.33}{37.78}{55.83} &
              \triplet{61.67}{52.38}{81.25} &
              \triplet{76.21}{57.79}{63.47} &
              \triplet{61.33}{46.54}{59.12} &
              \triplet{66.44}{49.54}{63.89} 
              \\
            \hline
            \vspace{1mm}\triplet{~}{CPT$+$}{~}&
              \triplet{\recall}{\F}{\precision}&
              \triplet{64.36}{59.63}{63.28} &
              \triplet{66.18}{60.38}{62.56} &
              \triplet{73.14}{54.72}{48.09} &
              \triplet{72.89}{59.91}{57.04} &
              \triplet{52.37}{58.22}{75.04} &
              \triplet{66.67}{64.59}{76.04} &
              \triplet{74.17}{63.10}{68.94} &
              \triplet{59.33}{56.45}{59.22} &
              \triplet{66.43}{60.20}{64.77} \\
            \hline
            \vspace{1mm}\triplet{~}{DG}{~}&
                \triplet{\recall}{\F}{\precision}&
                \triplet{66.40}{57.37}{57.33} &
                \triplet{62.29}{69.85}{75.00} &
                \triplet{71.78}{62.58}{61.03} &
                \triplet{68.79}{64.82}{72.73} &
                \triplet{72.90}{63.10}{56.25} &
                \triplet{71.66}{57.39}{49.45} &
                \triplet{72.11}{63.71}{61.55} &
                \triplet{60.63}{57.81}{60.23} &
                \triplet{66.85}{60.74}{60.43} \\
            \hline
            \vspace{1mm}\triplet{~}{LZ78}{~}&
                \triplet{\recall}{\F}{\precision}&
                \triplet{65.15}{56.89}{57.50} &
                \triplet{62.29}{69.85}{82.92} &
                \triplet{70.35}{59.31}{57.69} &
                \triplet{48.57}{48.18}{54.95} &
                \triplet{66.43}{66.67}{68.75} &
                \triplet{58.33}{57.48}{62.33} &
                \triplet{77.90}{62.88}{56.90} &
                \triplet{62.23}{58.72}{62.08} &
                \triplet{62.71}{58.75}{61.86} \\
            \hline
            \vspace{1mm}
            {\triplet{Markov}{Chain}{~}}    &
                \triplet{\recall}{\F}{\precision}&
                \triplet{63.16}{56.22}{57.40} &
                \triplet{100}{62.63}{47.42} &
                \triplet{70.61}{56.06}{51.48} &
                \triplet{63.64}{65.79}{65.91} &
                \triplet{58.33}{51.79}{47.50} &
                \triplet{64.17}{63.99}{77.50} &
                \triplet{72.11}{63.71}{61.55} &
                \triplet{60.84}{59.66}{64.30} &
                \triplet{68.92}{59.80}{59.39} \\
                \hline
            \vspace{1mm}\triplet{~}{TDAG}{~}&
                \triplet{\recall}{\F}{\precision}&
                \triplet{64.32}{55.57}{55.57} &
                \triplet{64.32}{67.59}{54.92} &
                \triplet{71.73}{59.09}{55.84} &
                \triplet{57.12}{50.69}{48.18} &
                \triplet{58.33}{56.94}{55.83} &
                \triplet{64.17}{63.99}{77.50} &
                \triplet{77.31}{63.40}{58.23} &
                \triplet{54.56}{54.56}{56.05} &
                \triplet{62.87}{57.90}{56.99}\\
            \hline
            \vspace{1mm}\triplet{~}{SubSeQ}{~}&
                \triplet{\recall}{\F}{\precision}&
                \triplet{31.98}{40.33}{60.80} &
                \triplet{28.96}{41.67}{81.25} &
                \triplet{31.29}{40.97}{66.14} &
                \triplet{41.97}{55.04}{87.12} &
                \triplet{38.67}{44.38}{58.33} &
                \triplet{48.33}{54.05}{65.00} &
                \triplet{32.29}{40.18}{60.20} &
                \triplet{34.06}{42.88}{63.27} &
                \triplet{34.80}{44.06}{68.92}\\
                \hline
            \vspace{1mm}\triplet{~}{\POIBert}{~}{\hspace{-5mm}} &
                \triplet{\recall}{\F}{\precision} &
                \triplet{58.87}{\textit{59.95}}{70.88} &
                \triplet{88.89}{62.63}{51.39} &
                \triplet{66.38}{59.75}{65.54} &
                \triplet{75.45}{62.70}{62.85} &
                \triplet{45.37}{45.37}{43.32} &
                \triplet{95.00}{62.96}{52.40} &
                \triplet{83.33}{63.92}{54.17} &
                \triplet{73.07}{55.92}{51.45} &
                \triplet{61.16}{62.32}{73.84} \\
            \hline
            \hline
            \vspace{1mm}\triplet{~}{\BTRec}{~}&
                \triplet{\recall}{\F}{\precision}
                &
                \triplet{59.40}{\textit{58.69}}{66.73}  &
                \triplet{64.44}{\textit{73.89}}{88.80} & 
                \triplet{64.28}{\textit{62.83}}{70.69}  &
                \triplet{72.73}{\textit{64.81}}{67.07}  &
                \triplet{72.92}{\textit{65.58}}{62.50}     &
                \triplet{69.44}{\textit{66.07}}{80.00}  &
                \triplet{63.60}{\textit{66.13}}{74.34}  &
                \triplet{66.61}{\textit{60.86}}{64.44}  &
                \triplet{65.01}{\textit{63.55}}{70.10} \\
            \hline
            \vspace{1mm}\triplet{~}{\bf \SBTRec}{~}&
                \triplet{\recall}{\F}{\precision}
                &
                \triplet{57.30}{\textbf{60.43}}{71.82}  &
                \triplet{71.11}{\textbf{75.56}}{82.22}  &
                \triplet{60.88}{\textbf{63.64}}{71.48}  &
                \triplet{69.70}{\textbf{67.55}}{74.75}  &
                \triplet{57.17}{\textbf{50.81}}{55.50}  &
                \triplet{80.00}{\textbf{66.67}}{71.10}  &
                \triplet{75.93}{\textbf{66.43}}{59.64}  &
                \triplet{59.45}{\textbf{62.71}}{76.15}  &
                \triplet{67.16}{\textbf{64.30}}{70.10} \\
            \hline
            \end{tabular}
            }
        }
}

\begin{document}
\maketitle

\begin{abstract}
    When traveling to an unfamiliar city for holidays, tourists often rely on guidebooks, travel websites, or recommendation systems to plan their daily itineraries and explore popular points of interest (\POIs). However, these approaches may lack optimization in terms of time feasibility, localities, and user preferences.
    In this paper, we propose the {\SBTRec}~algorithm: a \BERT-based Trajectory Recommendation with sentiment analysis, for recommending personalized sequences of \POIs~as itineraries.
    Considering the locations, sightseeing, and travel time between consecutive \POIs, our approach incorporates individual user preferences through the utilization of historical data.
    The key contributions of this work include analyzing users' check-ins and uploaded photos to understand the relationship between \POI~visits and distance.
    We introduce {\SBTRec}, which encompasses \emph{sentiment analysis} to improve recommendation accuracy by understanding users' preferences and satisfaction levels from reviews and comments about different {\POIs}.
    Our proposed algorithms are evaluated against other sequence prediction methods using datasets from 8 cities.
    The results demonstrate that {\SBTRec} achieves an average {\F}~score of 61.45\%, outperforming baseline algorithms.
    
    The paper further discusses the flexibility of the {\SBTRec} algorithm, its ability to adapt to different scenarios and cities without modification, and its potential for extension by incorporating additional information for more reliable predictions.
    Overall, {\SBTRec} provides personalized and relevant \POI~recommendations, enhancing tourists' overall trip experiences.
    Future work includes fine-tuning personalized embeddings for users, with evaluation of users' comments on \POIs,~to further enhance prediction accuracy.
\end{abstract}


\section{Introduction}
\if
    In the aftermath of the COVID-19 era, tourism continues to experience significant demand for a variety of reasons.
    Many people choose to go abroad, or travel to other countries due such as the ease of restrictions, escape from routine work, and relaxation.
    When individuals are getting ready for an international journey, a frequently employed method involves referring to guidebooks or online resources to structure their daily itineraries.
    On the other hand, they could choose to utilize tour recommendation systems that propose popular \POIs~based on their popularity, as showcased in prior research\cite{he2017category, lim2019tour}.
    Machine learning~(ML) has found diverse applications in fields like \emph{speech recognition} and \emph{machine translation}\cite{bert2019}.
    This paper focuses on the study of tour itinerary prediction using an ML technique.
    Specifically, Transformer models in ML have become the \emph{go-to solution}  for a range of natural language processing(\NLP)~tasks.
    This is attributed to their capacity to manage sequential data effectively and capture complex relationships\cite{bert2019}.
    Unlike other ML architectures such as recurrent neural networks and Long Short-Term Memory,~the Transformer model has the advantage of providing context for any position within the input sequence, enabling efficient parallel data processing.
    
    We focus on using techniques to tackle the challenge of predicting tour itineraries.
    Our innovative solution, named {\SBTRec}, uses a specific Transformer-based word embedding model designed to provide recommendations for a continuous sequence of \POIs.
    Our proposed algorithm is intended to assist tourists in proactively planning their travel itineraries based on their personal preferences. 
    Our methodology leverages historical data and reviews from Location-Based Social Networks(\LBSNs) of popular \POIs.
    Our algorithm takes into consideration a range of factors, encompassing locations, opportunities for sightseeing, and travel durations between successive \POIs.
    The complete workflow outlining our proposed model for predicting itineraries is outlined in Figure~\ref{figure:system}.
    In this paper, we present the following contributions:
\fi

In the \emph{post-COVID-19} era, there continues to be a significant demand for tourism, driven by various factors.
Many individuals opt for international travel due to factors such as the relaxation of travel restrictions, the desire to escape from their daily work routines, and the need for leisure.
When people prepare for international trips, they commonly turn to guidebooks or online resources to plan their daily schedules.
Alternatively, they can utilize tour recommendation systems that suggest popular points of interest~(\POIs) based on their popularity, as demonstrated in prior research\cite{he2017category, lim2019tour}.
Machine learning~(ML) has found diverse applications in various fields, including speech recognition and machine translation\cite{bert2019}.
This paper focuses on exploring ML~techniques for predicting tour itineraries.
In particular, \emph{Transformer} models in ML have emerged as the preferred solution for numerous natural language processing~(\NLP) tasks with their high accuracy in handling sequential data effectively and capturing intricate relationships\cite{bert}.
Unlike other ML architectures like Recurrent Neural Networks and Long Short-Term Memory, the Transformer model possesses the advantage of offering context for any position within the input sequence, facilitating efficient parallel data processing.

Modern technology enables tourists to have reliable high-speed internet access: this suggests that tourists can now easily connect to the internet with their smartphones or tablets, even when they are traveling.
This allows them to access information and services that they need, such as \POI~recommendations, maps, and transportation schedules.
As a result, tourists often seek new \POIs~for sightseeing ideas: When planning an itinerary trip, tourists often want to find new and interesting places to visit.
They can use their smartphones or tablets to search for~\POIs, read reviews, and get directions.
In this paper, our focus is on employing techniques to address the challenge of predicting tour itineraries problem.
Our innovative solution, namely {\SBTRec}, leverages a specific Transformer-based word embedding model designed to provide recommendations for a continuous sequence of {\POIs}.
The objective of our proposed algorithm is to aid tourists in \emph{proactively} planning their travel itineraries based on data on past users' trajectories and individuals' preferences in selecting \POIs.
Our approach incorporates historical data and \POI~reviews from  Location-Based Social Networks~({\LBSN}s) related to popular \POIs.
Our algorithm takes into account multiple factors, including geographical locations, sightseeing opportunities, and travel durations between consecutive \POIs.

In this paper, we present the following contributions:
\begin{itemize}
    \item 
        We propose {\SBTRec}, a \BERT~embedding model that recommends {\POIs} as an itinerary based on the check-in records from users’ past trajectories, such as their timed records and \POIs~metadata such as time/GPS locations.
    \item 
        To capture users' travel preferences and patterns of~{\POI}, a selection that is not effectively represented in existing models, we propose a transformer-based approach that analyzes users' past visits by training on a large dataset of photos and their timestamp distribution during their visits to~{\POIs}.
        This model is trained to uncover these underlying patterns and preferences, enabling it to make more personalized and effective~\POI~recommendations.
    \item 
        The proposed {\SBTRec} algorithm integrates sentiment analysis into the prediction algorithm for \POI~itineraries, leading to improved accuracy of itinerary predictions..
    \item 
        We propose the addition of {\NextPop}~gate to fine-tune the {\MLM} prediction task of the \POI-{\BERT} model.
        The {\NextPop}~gate aggregates numeric values of the input data of the problem, such as the total number of photos uploaded to 
        {\LBSN}~and visitors' reviews that are usually presented in the form of human language.
        This information can lead to more accurate predictions.
    \item 
        We assessed the performance of different cities in our experiments. The results from our experiments, as presented in Section~\ref{section_4}, demonstrate the consistent and reliable ability of our proposed algorithm to predict itineraries, achieving an average \F-score accuracy of 61.48\% across the 8 cities in our datasets.
    \item 
        Finally, our proposed algorithm has the advantage of adapting to different scenarios~(cities/datasets) without tuning and modification. Furthermore, we observed a performance increase of up to 12.93\% in our {\Glas}~dataset compared to other implementations~(from 64.81\% to 67.55\% measured in averaged \F~score.)
\end{itemize}

The rest of this paper is organized as follows:
Section~\ref{section_related_work} provides background on tour recommendations and discusses the state-of-the-art approaches to the itinerary prediction problem.
In Section~\ref{section_formulation}, we formally define the tour itinerary prediction problem and introduce the notations used in our proposed solution.
Section~\ref{section_experiments} outlines our experiment framework and presents the baseline algorithms used for evaluating the effectiveness of our solution.
Finally, We conclude our paper in Section~\ref{section_conclusion}, where we discuss the implications of our findings and suggest directions for future research.

\section{Preliminaries}
    \label{section_2}
    \label{section_related_work}
    \label{RELATED_TOUR_RECOMENDATION}

    In this Section, we begin by evaluating the current solutions available for producing recommendations for \POIs~and predicting sequences in Section~\ref{RELATED_SEQ_PREDICTION}. 
    Moving on to Section~\ref{section_2_2}, we assess the latest cutting-edge solutions used for generating tour recommendations.
    In Section~\ref{section_2_3}, we delve into the solutions related to sentiment analysis and examine how they are applied in sequence prediction in practice.
    In Section~\ref{section_2_3}, we explain some solutions related to sentiment analysis and examine how they are applied in sequence prediction in practice.

    \iftrue
    \begin{table}[t]
        \centering
        \caption{Notation used in the paper}
        {\notationTab}
        \label{tbl:notations}
    \end{table}
    \fi

    \subsection{Sequence Prediction}{
        \label{section_2_1}
        \label{RELATED_SEQ_PREDICTION}
        Sequence prediction is a foundational challenge within machine learning, focused on predicting the next item in a sequence based on previously observed ones\cite{chen2022sequential}.
        This problem uniquely considers item order, as seen in applications like time-series forecasting and product recommendations\cite{chen2022sequential}.
        For tour recommendation, sequence prediction is adapted to anticipate a traveler's next~\POI~visit.
        By treating a user's itinerary as a sequence of locations, this approach aims to predict the following \POI, accounting for locality.
        Existing models integrate {\WordVec}~techniques such as {\SkipGram}, \CBOW}, and {\LSTM}~networks to represent \POIs~as words\cite{poibertbigdata22,word2vec_rec_2020,li2022transformerbased}.
        Other works further incorporated the \emph{spatio-temporal} information into the recommendation system\cite{Spatiotemporal_LSTM_2021}.

    \subsection{Tour Recommendation \& \POI~Embedding}{
        \label{section_2_2}
        The research covers the \emph{next-location prediction}\cite{sohrabi2020greedy,lim2019tour} and \emph{itinerary recommendation/ planning}\cite{lim2019tour}.
        Specialized algorithms leverage check-in data from location-based social networks~(\LBSN) to suggest tailored itineraries, considering user preferences and similar patterns.
        ML-based algorithms  recommend \POIs~based on past check-ins, considering locational data to predict the next \POI\cite{mustsee_poi_2018,Spatiotemporal_LSTM_2021}.
        The {\POIBert}~model enhances prediction using a special encoded of \BERT~language model from user trajectories\cite{poibertbigdata22}, although personal preferences are limited in their embedding model.

        The \emph{next~location prediction} challenge pertains to the process of identifying the next \POIs~a tourist is more likely to visit, while also taking into account patterns observed in the activities of other travelers\cite{Zhuang2017}.
        Personalized tour recommendations have been crafted by leveraging check-in data sourced from~{\LBSN}s.
        They provide detailed and updated information from users of {\LBSN}s, such as check-in information with time sensitive GPS locations.
        This check-in data encompasses valuable details, including photos and embedded metadata for analysis of \POI~recommendation.
        By analyzing this data, specialized recommendation algorithms can be tailored to align with the unique interests and preferences of individual users.
        In prior research on \POI-recommendations, the emphasis has predominantly centered on suggesting popular \POIs,~factoring in considerations such as waiting times and ratings\cite{cai2018itinerary, geolocation2023, gnn_preference_2020}.
        Additionally, other works also explored the use of \emph{geo-tagged} photos shared on {\LBSN}~to collect valuable information about a wide range of {\POIs}\cite{geolocation2023}.

        Different ML~algorithms have been proposed to recommend popular \POIs~based on past check-in data and trajectories\cite{halder2022poi}.
        These methods use locational data collected to predict the next~{\POI} such that users are most  likely  to be at the check-in location\cite{Zhuang2017}.
        However, such a method considered a limited number of factors and did not provide the full detailed itinerary.
        The {\POIBert}~model is first proposed by considering the check-ins and duration of users' trajectories as input to the \BERT~language model for training of the {\POI}-prediction task\cite{durationrectour22}; 
        the algorithm is used to predict itineraries by regarding:
          i) users' trajectories as~\emph{sentences} and
          ii) travels visit to \POIs~as \emph{words} into the training of~\BERT~model.
        The {\POIBert}~algorithm recommends an itinerary by iteratively predicting the next \POI~(as the next `word')~to visit using the~\MLM~prediction model.
        However, their recommendation takes into account limited \emph{user's preferences} by considering a selection of \emph{initial} and \emph{destination}~{\POIs} when planning users' daily itineraries.

        {\CorpusGeneration}

    } 

    \subsection{\BERT~classification}{
        \label{section_2_3}
        Bidirectional Encoder Representations from Transformers~(\BERT) classification is first used as an \NLP~technique  for solving text classification tasks\cite{bert}. 
        It is now a \emph{de~facto} model for pre-trained language modeling to understand the contextual relationships between words\cite{ImprovingBert2021}.
        {\BERT}~classification is shown to have impressive performance in many \NLP~tasks due to its ability to capture contextual information and transfer knowledge.
        The high performance of the {\BERT}~model is achieved by training using the~\emph{Masked Language Model}~({\MLM})~and \emph{Next Sentence Prediction}~({\NSP}) algorithms.
        The results of the two algorithms are then combined using a \emph{loss function}.
        In \MLM, the \BERT~model is trained to predict randomly \emph{masked} words based on the surrounding \emph{context}.
        On the other hand, {\NSP} training aims to determine whether two sentences appear consecutively in a given \emph{context}.
        {\BERT}~model regards \emph{corpus} as text tokens which may include numeric values passed to {\BERT} for prediction task.
        Attempts have been made to include numerical values with text values in the \BERT~prediction.
        These approaches, however, only use the \BERT~layer for the prediction of  \emph{textual information}; and later combined with the numerical values to produce a \emph{multi-modal feature} for any downstream tasks\cite{nlp_know_numbers_2019,Multimodal_Toolkit_2021}. These numeric values are not interpreted in the \BERT~training task.     

        Previous works suggest that itinerary prediction can be solved by using specific language model by training a {\BERT}~language model using a corpus(training samples) consisting of users' trajectories in the form of \emph{sentences}, where every \emph{word} represents check-in information, described in Algorithm~\ref{alg:CorpusGeneration}.
        Hence the algorithm outputs training samples of size~$O(k \cdot N)$ for the downstream classification tasks, where $N$ represents the size of users' trajectories and $k$ denotes the number of \POIs~of the longest path of trajectory.
        In order to prioritize recommending itineraries, the {\BTRec} model is proposed by incorporating users' demographic information into the training process~\cite{BTrec_2023}.
        Due to the focus on the training of long trajectories, short trajectories are less well-represented\cite{BTrec_2023}.
        Therefore, in Section~\ref{section_formulation}, we propose fine-tuning the prediction model by incorporating the \NextPop gate to forecast the next POI to be included in the partial solution.

    }

    \subsection{Sentence-\BERT~Embedding}{

        \label{section_2_4}
        Sentence embedding is a technique in \NLP~that represents a sentence as a vector of numeric numbers, with the aim of
        capturing the semantic meaning of the sentence, so that they can be used for other downstream a variety of tasks, such as sentiment analysis and text classification.
        A common method to create sentence embeddings is to use an artificial neural network to learn a mapping from \emph{sentences} to \emph{vectors}, by training on a \emph{corpus} of text, to associate each sentence with a vector such that it captures its meaning.
        Sentence embeddings have been used as an effective classifier  for many \NLP~tasks.
        For example, they have been used to improve the accuracy of sentiment analysis models and to develop more effective text classifiers.
        Sentence embedding has been shown to be effective in comparing the similarity between sentences and learning the \emph{semantic relationships} between words and sentences\cite{reimers2019sentencebert}.

        Previous studies on {\POI} itinerary prediction have used {\POI} embedding to only consider the popularity of {\POIs} based on the frequency of visits.
        However, they do not consider the impressions of visitors after visiting {\POIs}, which they may share on popular LBSNs
        for other potential travelers.
        In this section, we discuss how users' reviews can impact others' decisions about which {\POIs} to visit using sentiment analysis. {\BERT}~is ineffective for tasks  involving semantic search in sentences, which can lead to significant training overhead\cite{sbert2020}.
        S-{\BERT} is designed to overcome this problem by learning "meaningful representations" of individual sentences, simplifying the heavy computational load of similarity comparisons.
        It provides a \emph{lightweight}~{\BERT}~extension based on the goal of maximizing mutual information.
        Additionally, a typical S-{\BERT} embedding is a vector of low dimension, which can be easily compared against other embeddings using simple numerical operations; it also has the advantage of using fewer system resources during the process of training.

        \iftrue
            \begin{figure*}[h]
                \centering
                {\includegraphics[width=1.02\textwidth,
                  trim=0.0cm 0.0cm 1.6cm 0.0cm,
                  clip
                  ]{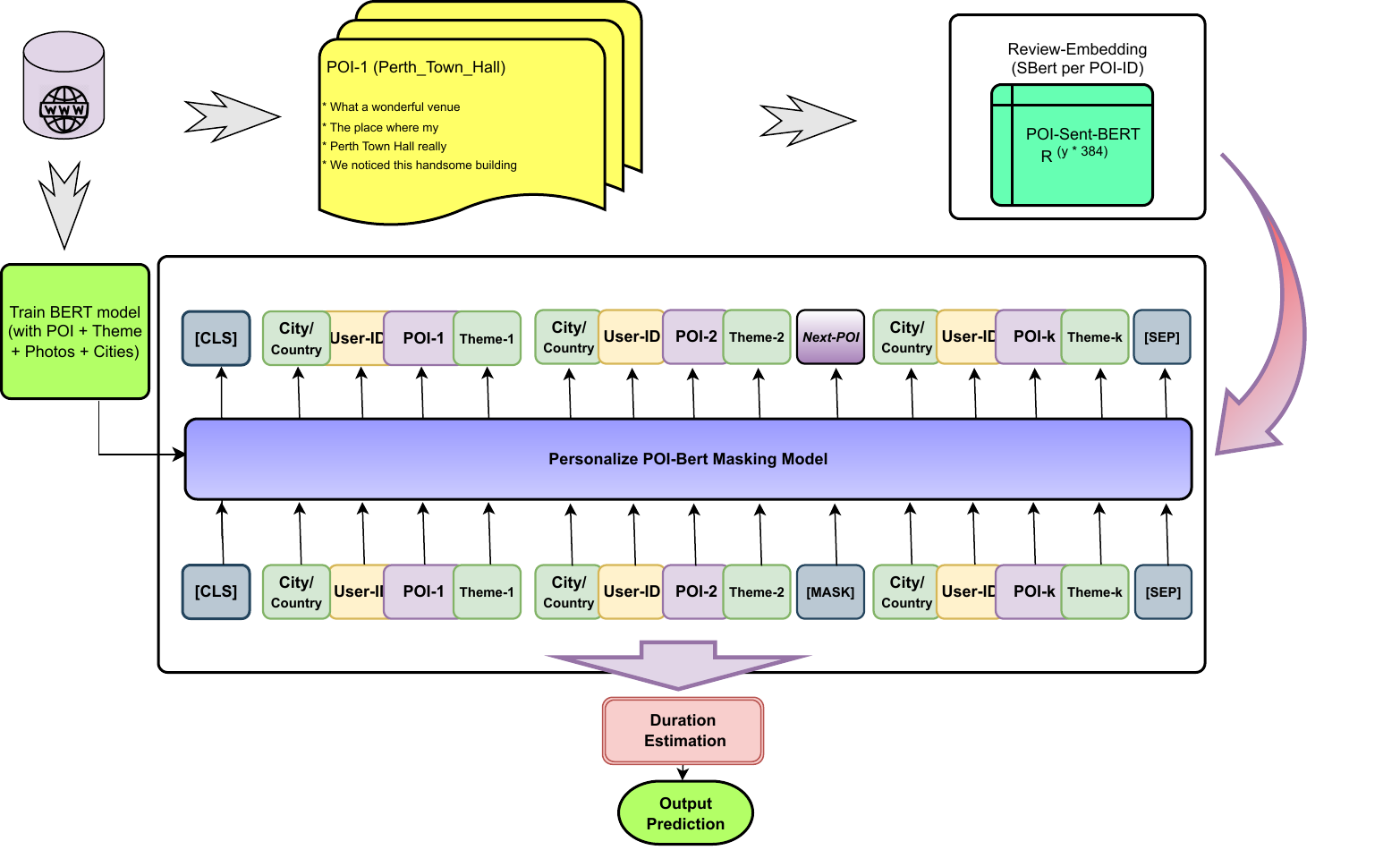}}
                \caption{Overall \SBTRec~system workflow of itinerary prediction using users' comments and trajectories  }
                \label{fig:workflow}
            \end{figure*}
        \fi
            
        In our studies of \POI~embedding, previous solutions only considered the {\POI}~popularity by the rate of visits to these \POIs.~However, they did not consider visitors' impressions after visiting the \POIs,~which they may share on popular LBSNs for other potential travelers.
        In this section, we discuss how users' reviews can impact others' decisions in choosing \POIs~to visit using sentiment analysis.
        {\BERT}~has been shown to be inefficient for tasks involving semantic search in sentences, which can lead to significant training overheads\cite{sbert2020}.
        S-{\BERT}~is used to solve the problem by learning `meaningful representations' of individual sentences, simplifying the heavy computing load of similarity comparisons.
        It thus provides a `lightweight'~{BERT}~extension based on the goal of mutual information maximization.
        Moreover, a typical S-{Bert}-embedding produced are vectors of low dimension, which can be easily compared against other embeddings using simple numeric operations.

    }

\section{Problem Formulation and Algorithms}
    \label{section_formulation}
    \label{section_3}

    This section formally presents the problem of tour itinerary recommendation in this study.
    To simplify our discussion and presentation, we will be using symbols and terms that are summarized in Tab.\ref{tbl:notations}).
    Consider a group of users, $U$, who have uploaded photos to a~\LBSN.
    These photos were taken at various \POIs~while visiting a city.
    There are a total of $|P|$~\POIs, and tourists checked in at several of them, taking $|C_i^{u}|$~photos with timestamps during their travels.
    These check-in records represent a list of check-ins at {\POI}-$pi$, with each record containing the timestamps of photos taken and posted on the \LBSN.
    The list of check-in records at \POI-${p_i}$ forms a sequence of $\{(p^u_1,c^{u}),(p^u_2,C^{u})$, where $(p^u_k,C^{u})$ tuples denoted as the set of check-in records $C^u = [(c^u_1, t^u_1), (c^u_2, t^u_2), ..., (c^u_k, t^u_k)]$, where $\forall \in {POI}_{j}$ represents the timestamps of the photos taken and posted to the location-based social network({\LBSN}).
    The main focus of this research is to propose a \emph{personalized} itinerary  of Points of Interest~(\POIs) that users are more  likely  to visit.
    The itinerary recommendations are based on users' past trajectories gathered from~\LBSN. 
    The paper considers the starting and ending \POIs, denoted as $p_u$ and $p_v$ respectively, and utilizes the photo and check-in data available at the starting \POI~($p_u$).

    \begin{paragraph}{Sentiment Analysis via {S-{\BERT}} Embedding}  
        Sentiment analysis is a well-explored field. Various users' reviews or comments posted on \LBSN~are significant resources for potential tourists to gather insights before their visits.
        As part of our system, we introduce a component aimed at analyzing these comments and investigating how they \emph{impact} users' decision-making when selecting \POIs~to visit.
        Previous research on \POI~recommendation has often relied on metrics like the total number of visitors to gauge \POI~popularity.
        However, some visitors may capture a few photos, while others may take more photos at some particular \POIs;
        they may also opt to share \emph{negative reviews} as an expression of dissatisfaction with particular \POIs.
        To address this, our algorithm assesses the level of satisfaction experienced after visiting \POIs~by assessing the \emph{photo counts} and the duration of staying at the \POIs~throughout their itineraries.
        At the same time, we also analyzed the \emph{top reviews} and evaluated the impact.
        By scrutinizing the \emph{sentiments} conveyed in different users' reviews about \POIs,~tourists can gain deeper insights into their preferences and satisfaction levels when deciding their next~\POIs. Fig.~\ref{fig:Sbert_comments} shows some examples of users' reviews on two \POIs~in our Perth dataset.
        We conjecture that positive reviews will lead to more tourists.
        To achieve this, we propose using the lightweight \SBert~embedding model to map a user comment to a representation that  `\emph{maximizes global textual information}'.
        By mapping each user's comment as an {\SBert}~embedding,
        we intend to model users' comments at a comparable representation of users' rating of a \POI,~which can then be \emph{normalized} and evaluated numerically, for each~\POI~in the city of interest.
        They are then aggregated as a group of users' embeddings for measuring users' sentiment to a~\POI.
        
        This information empowers tourists to personalize their itinerary recommendations more effectively, ensuring that the recommended \POIs~align closely with their interests and preferences.
        This refinement enhances our prediction algorithm by considering popular~\POIs~as expressed as languages.
        This refinement enhances our prediction algorithm in the dimension of \POI~popularity.
    \end{paragraph}

    \begin{equation}
\includegraphics[width=12.9cm, trim=2.02cm 7.6cm 7.00cm 18.6cm,clip]{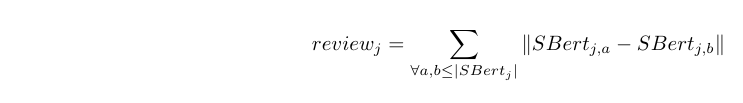}      \label{NextPopEqu}
    \end{equation}  
        
     Equation~\ref{NextPopEqu} above measures the normalized  distance between any \emph{pair} of comments in the dimension of \SBert~for any $\POI_j~\in \POIs$.
     Intuitively, positive reviews/comments posted to some {\POI} with positive reviews are similar to each other in the dimesnions of its \SBert~embedding. Therefore, such \POIs~with  positive comments will have a higher weighting using the division operation in the \NextPop~gate.

    \begin{table}
        \centering
        \caption{Examples of users' comments in Perth dataset showing  \POIs}
        \includegraphics[width=10.8cm, trim=8.02cm 7.0cm 7.20cm 1.6cm,clip]{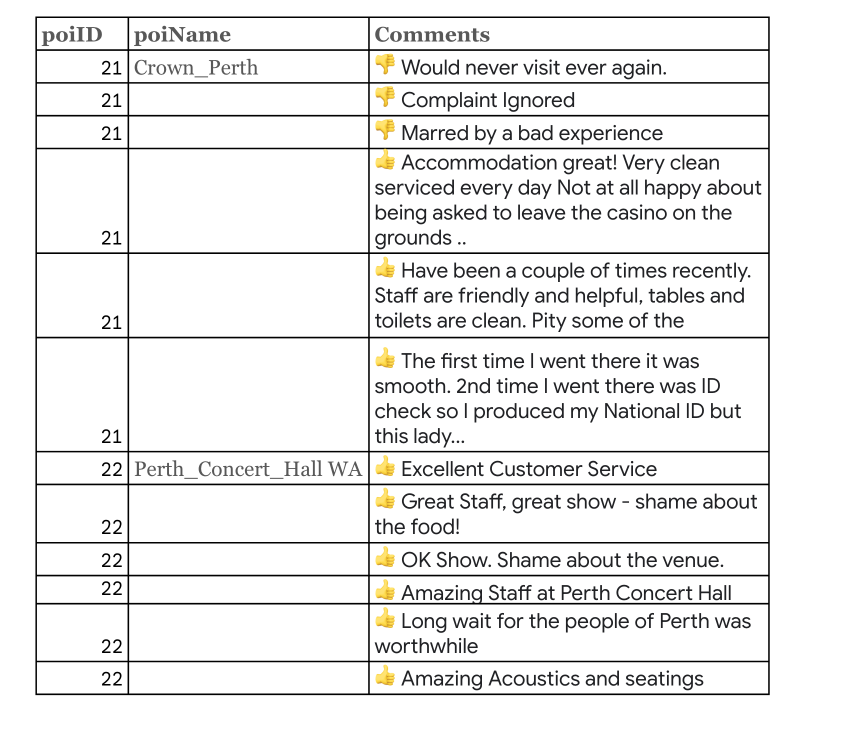}
        \label{fig:Sbert_comments}
    \end{table}

    \begin{paragraph}{Effect on Multiple Photos Uploaded at the Same \POI}
        It is important to note that not all visitors take many photos at \POIs, and some only have a few photos.
        To address this, our proposed model assesses satisfaction by analyzing the number of photos being uploaded, the duration of stay at \POIs,~and the top reviews from travel websites.
        We estimate the effects of uploaded photos and their influence on different travelers in choosing their next~\POI. We conjectured that tourists are attracted to visit \POIs~with more photos posted online.
        Our itinerary algorithm considered the fact that the dataset the model built upon is biased towards \POIs~with more photos.
        Hence we made adjustments to our prediction model by \emph{dividing} the estimated photo count~($\hat{F_i}$) of a $\POI_i$ instead. $\forall i \in \POIs$.
        We employ a similar treatment of estimating duration to estimate the \emph{expected photo count} of a \POI~ by finding the confidence~interval of the photo count uploaded for any \POI.
        Hence, estimate the \emph{expected} photo counts by calculating the 90\%-\emph{confidence interval} from our dataset using a statistical method of \emph{bootstrapping}, similar to computing the duration of visits to \POIs\cite{poibertbigdata22}.
           %
        
        
    \end{paragraph}

    \begin{paragraph}
    {\NextPop : A Refinement Gate to Next \POI~Prediction}
      We propose to improve the performance of the \POI~
      recommendation system by taking into account a few important factors when selecting the next \POI~to be inserted in the proposed itinerary, such as:
        \begin{itemize}
            \item \POI~prediction:
              The \MLM~prediction algorithm is a type of language model that is trained to predict the next word in a sequence.
              In the context of \POI~recommendation, the \MLM~prediction algorithm can be used to predict the next~\POI~in a user's itinerary.
              This is done by taking into account the context of the previous \POIs~in the itinerary.
            \item Sentiment analysis:
              Sentiment analysis is the process of determining the sentiment of a piece of text.
              In the context of \POI~recommendation, sentiment analysis can be used to determine whether users have a positive or negative opinion of a~{\POI}.
              This can be done by analyzing the text of     users' comments about the~\POI.
            \item Photo-popularity:
              it is a  measure of the number of photos that have been taken at a POI.
              This can be used to estimate the \emph{popularity} of a~{\POI}. {\POIs} that have been photographed more often are likely to be more popular than \POIs~that have been photographed less often.
        \end{itemize}
      By taking into account these factors, a {\POI} recommendation system can be more effective at recommending {\POIs} that are relevant, diverse, and enjoyable for the user.
      The refinement gate then uses these factors to make a more informed decision about which \POI~to recommend. This results in a more \emph{accurate} and \emph{personalized} \POI~recommendation system.

      The \MLM~prediction is made by the original {\BERT}~model, which is a pre-trained language model.
      The {\BERT}~model is trained on a massive dataset of text and code, and it can be used to make predictions about the next \POI~in an itinerary.
      The sentiment analysis is performed on users’ comments about the \POI.
      This analysis helps to determine whether the \POI~is generally liked or disliked by users.
      The photo~popularity is the number of photos that have been taken at the \POI.
      This metric is used to measure the interest of tourists in the \POI.
      The refinement gate then uses these three factors to make a decision about which \POI~to recommend.
      The gate considers the results of the \MLM~prediction, the sentiment analysis, and the photo-popularity.
      The gate then outputs a score for each \POI, and the \POI~with the highest score is recommended.
    \end{paragraph}
    An analogy for applying the {\NextPop}~gate is tourists may seek advice from a few {\LBSN}s while also considering the \emph{popularity} of photos and comments before they make their final decision on choosing a \POI~to visit.

    \begin{paragraph}{Itinerary Prediction of \SBTRec~Algorithm }
    Prediction of a \POI-itinerary generally takes inputs as the \emph{source} and \emph{destination}~\POIs, $p_u$ and $p_v$, respectively, and the total time budget of the itinerary. 
    As described in Algorithm~\ref{alg:SBTRec}, the prediction algorithm starts by asking from the training data~set for $u'$, the \emph{closest} \emph{reference}~user that is associated with \POIs~$p_u$ and $p_v$.
    This is achieved by solving a series of~\MLM~problems as detailed in the first two lines of the Algorithm~ref{alg:SBTRec},  so as to maximize the score of the {\MLM} query.
    The rest of the prediction algorithm is to iteratively find an \emph{unvisited} \POI~and insert it into the predicted itinerary while maximizing the prediction score in every iteration while the time budget is not exhausted.

    {\SBTRecAlgoOneColumn}

    \end{paragraph}

    \section{ Experiments and Results }
        \label{section_experiments}
        \label{section_4}
    
        The data set we used in our experiments is a collection of photos uploaded to the Flickr platform
          \footnote{Source code is available at: 
          \hyperref[https://nxh912.github.io/SBTRec_BigData23/]
          {\texttt{http://tinyurl.com/595mbzzd}}}. 
        The photos capture the trajectories of 5,654 users from eight popular cities.
        The photos are labeled with metadata, such as the date, time, and GPS location.
        We sorted the photos in the data set by time and then mapped them to the relevant \POIs~using their GPS locations.
        We then reconstructed the travel trajectories of all users who visited at least 3 {\POIs}.
        This process generated sequences of time-sensitive \POI~IDs that represent the users' trajectories over time.
        We utilize the S-{\BERT}~embedding for sentiment analysis\footnote{ 
            \scriptsize These comments are trained using a \SBert~language model "\texttt{\href{https://huggingface.co/sentence-transformers/all-MiniLM-L6-v2}{sentence-transformers/all-MiniLM-L6-v222}}"}%
        in our \SBert~model  prepared, by through analyzing  users' comments%
          \footnote{ 
            \scriptsize Tourists' top~20 comments are collected from: \href{https://www.tripadvisor.com/}{Tripadvisor.com}
          }
        posted in \LBSN\cite{sbert2020,allMiniLM}.

        \subsection{Datasets}
        The data sets consist of approximately 170K photos or check-in records collected from 6681 users in eight popular cities\cite{poibertbigdata22}. 
        %
          Our data sets have been divided into three distinct sets: Training, Validation, and Testing data sets.
          Initially, we sorted all photos according to their Trajectory-IDs based on their \emph{last check-in times} in ascending order.
          To generate the Training Data set, we set aside the first 70\% of trajectories based on their associated photographs.
          The subsequent 20\% of trajectories were assigned to the \emph{validation} set, while the remaining data was assigned to the \emph{testing data set}.
          This method of segregating the data helps to prevent the issue of a  trajectory being present in multiple data sets.
        
        \subsection{Baseline Algorithms for Performance Comparison}
             The following baseline algorithms are used for performance comparison:
            \begin{itemize}
                \item {\spmf}~algorithms -
                this software package encompasses a collection of algorithms designed to forecast the subsequent symbol in a sequence using a set of training sequences, such as:
                    {\CPT}, {\CPTplus},~{\TDAG}, {Markiv Chain} and {Directed Graph}%
                    \cite{CPT2013, CPTplus2015, DG1996, TDAG, PPM1984, AKOM1999}.
                \item{\SuBSeq}: 
                  the algorithm employs compressed data structures to efficiently store and manipulate the subsequently as a \emph{``Succinct Wavelet Tree''} data structure\cite{succinctBWT_2019}.
                \item {\POIBert}:
                   it relies on the {\MLM} algorithm in a fine-tuned~{\BERT}~model to generate predictions in choosing \POIs\cite{poibertbigdata22}. Additionally, it employs \emph{bootstrapping} to gauge the lengths of \POI~visits by estimating the duration of visits in the \POIs.
               \item {\PerPoiBert}:
                   this algorithm enhances the {\BERT}~embedding~model by training the customized embedding, using a curated \emph{corpus} incorporating users' demographic information into the {\POIBert}~model~\cite{BTrec_2023}.
            \end{itemize}
            
            Some baseline algorithms in \spmf~package predict the \emph{next token}~(as a \POI), our sequence prediction task encompasses the iterative prediction of further tokens (as~\POIs)~until the user-defined time limit is attained.
            To evaluate the efficiency of both our proposed algorithms and the baselines, we carried out all experiments in a uniform manner as described in Section~\ref{accuracy}.
            In these experiments, the algorithms utilized identical datasets for \emph{training}, \emph{validation}, and \emph{testing} purposes.
    
        \subsection{Performance of Algorithms}
        
            \label{accuracy}
            We performed experiments in eight cities from the Flickr dataset.
            We considered all trajectories from users as sequences of \POIs~(\emph{corpus}).
            To assess the performance of our models, we trained various sequence prediction models with different hyper-parameters.
            The accuracy of these models was evaluated using the Validation and Test sets:
              for each trajectory in the dataset, referred to as the \emph{history-list}.
            We considered the first and last \POIs~as the \emph{source} and \emph{destination} \POIs~of the \emph{query} itinerary;
              we also regard the time allocated for the \emph{query} as the time difference between the first and last photos of each trajectory. 
              We then use our prediction models to recommend the \emph{intermediate} \POIs~of the trajectory within a specified time frame.
            We conducted experiments in eight cities using the Flickr dataset. We analyzed user trajectories when they visited at least 3 {\POIs} in the training set. These trajectories were treated as sequences of {\POIs}, forming a \emph{corpus}. 
            To gauge the effectiveness of our models, we trained various sequence prediction models with different hyper-parameters. The accuracy of these models was assessed using Validation and Test sets. For each trajectory in the dataset, referred to as the history-list, we identified the first and last \POIs~as the source and destination \POIs~for the itinerary prediction query. The time allocated for the query was determined as the time difference between the first and last photos of each trajectory.
            We evaluated the performance of the {\SBTRec}~prediction algorithm by using the precision~($\mathcal{T}_{\mathcal{P}}$), recall~($\mathcal{T}_{\mathcal{R}}$), and~\F~scores, comparing the recommended \POI~trajectory with the actual \POI-\emph{path} using the following evaluation metrics:
            \iftrue
                \noindent  Let $S_p$ be the predicted sequence of \POIs~from the algorithm, and $S_h$~be the actual sequence from the trajectories, we evaluate our algorithms based on: 
                \begin{itemize}
                    \item $\mathcal{T}_{\mathcal{R}}(S_h,S_p)$ = $ \frac{|S_h \cap S_p|}  {|S_p|}$, 
                    \item $\mathcal{T}_{\mathcal{P}}(S_h,S_p) = \frac{|S_h \cap S_p|}{|S_h|}$, and,
                    \item $\mathcal{F}_1\_score(S_h,S_p) =  \frac{ 2 \cdot  \mathcal{T}_{\mathcal{R}}(\bullet) \cdot \mathcal{T}_{\mathcal{P}}(\bullet)}{ \mathcal{T}_{\mathcal{R}}(\bullet) + \mathcal{T}_{\mathcal{P}}(\bullet) }$
                \end{itemize}
            \fi
            
            \begin{paragraph}{Tuning of hyper-parameters}
                In the pursuit of identifying the most suitable \emph{hyper-parameters} for our experiments, we conducted training on the \SBTRec~models with varying \emph{epochs}, spanning from 1 to 60, utilizing the \emph{Training} dataset.
                Subsequently, these models were employed to predict itineraries within our \emph{validation} dataset.
                The model that demonstrated the highest average \F~score of predictions across the             \emph{validation} dataset was chosen. Finally, the accuracy of prediction was reported using the selected model to generate recommendations for the \emph{test} dataset.
                We also note that~algorithms in {\spmf} package have no hyper-parameters for tuning\cite{spmf2017}. 
            \end{paragraph}

        \subsection{Experimental Results }
            We assessed the effectiveness of our proposed algorithms in various cities by constructing travel histories based on the chronological ordering of photos. 
            The accuracy of the predicted itineraries was compared in terms of average \F~scores in Table~\ref{tbl:ExptTab}.
            To compare the results of our proposed model with other baseline algorithms,
            we \emph{reproduce} some experimental results of the baseline algorithms below. 
            This allows us to conduct a complete analysis of our proposed algorithm,
            which is based on past work on trajectory recommendation\cite{poibertbigdata22, BTrec_2023}.
            Overall, the experimental results in Table~\ref{tbl:ExptTab} suggest that the {\SBTRec}~itinerary prediction algorithm achieves a significant improvement in the itinerary prediction tasks.
            Our proposed {\SBTRec} algorithm achieved 64.00\% on average, which significantly outperforms the~{\POIBert}~algorithm with an average {\F~score} of 56.86\%, on average.
    
            \begin{table*}[!th]
                \centering \caption{\scriptsize{Average~Recall({${\mathcal{R}}$})/{\F}/{Precision(${\mathcal{P}}$)}~scores of~prediction algorithms in Test~datasets~(\%)}}
                
                {\ExptTab}
    
                \label{tbl:ExptTab}
            \end{table*}
    
            Our proposed {\SBTRec} algorithm can generally recommend tour trajectories that are more \emph{personalized} to users' preferences and interests to them, compared to the actual trajectories.
            The {\SBTRec} algorithm further enhances the prediction of the \POI~itineraries by incorporating popular {\POIs} by their photo count into the embedding model.
            In all experiments in eight cities, {\POI} trajectory predictions using the {\SBTRec} algorithm can generally predict itineraries with an average {\F}-score of 66.53\%.
            Our proposed {\SBTRec} algorithm outperforms other baseline algorithms in predicting tour itineraries.
            On average, without tuning of hyper-parameters, the {\SBTRec} algorithm can \emph{generally} predict itineraries with an average {\F}-score of 58.05\% in all datasets and hyper-parameters,
            while the next~best algorithm (\PerPoiBert~algorithm) only predicts itineraries with an average {\F}-score of about 56.45\%.
    
        
    Our proposed algorithm, {\SBTRec},~outperforms baseline prediction algorithms in terms of prediction accuracy. 
    While baseline algorithms like {\CPT}~and {\SuBSeq}~rely solely on sequences of words representing past {\POI} trajectories, our transformer-based architecture effectively leverages the relationships between {\POIs} and their corresponding themes, incorporating individual users' demographic information for enhanced prediction.
    Among other transformer-based baseline algorithms, the {\POIBert}~and {\BTRec} demonstrate promising performance.
    Furthermore, our {\SBTRec} algorithm achieves superior prediction accuracy by incorporating the {\NextPop}~gate into the transformer-based prediction model by identifying \emph{popular} {\POIs}~with positive reviews from \LBSN.

\section{Conclusion}
    \label{section_conclusion}
    \label{section_5}
    
    In this paper, we present {\SBTRec}, a novel method aimed at assisting tourists in planning an optimized travel itinerary.
    The system recommends a sequence of \POIs~by taking into account factors like location, time limitations, and individual preferences in selecting \POIs.
    Our method involves creating and training a language model based on {\BERT}~with a novel {\NextPop}~gate,~which is fine-tuned to enhance the recommendation process of finding a new \POI~to visit. This approach employs training, validation, and test datasets to ensure accurate and tailored suggestions.
We utilize the {\POI} {\BERT}-based classification, our objective is to offer tourists a more in~depth and \emph{context-aware} approach to planning their itineraries.
    Furthermore, we have developed the {\NextPop~gate}, which allows our {\BERT}~model to undermine tourists' decision preferences in selecting a \POI~for a visit with the consideration of external factors, such as influence from comments and photo counts contributed by past tourists in \LBSN.
    
    Our algorithm involves analyzing the \emph{source} and \emph{destination} \POIs, it can accurately determine users' preferences for selecting intermediate \POIs~they are more likely to visit during their site-seeing.
    Our {\SBTRec}~prediction algorithm uses a statistical method of finding the duration of visits from past trajectories with a high confidence level.
    To ensure the reliability of our model, we conducted extensive experiments to analyze the performance of our algorithm. 
    It show cases the effectiveness in predicting relevant \POIs~based on \emph{recall}, \emph{precision}, and \emph{\F}~scores.
    Furthermore, the adaptability of our proposed algorithm to diverse scenarios~(various cities and different \POI~themes/categories) was demonstrated through experiments conducted across eight cities.
    Our approach, which factors in check-in frequencies, and \POI~locations with users' feedback with sentiment analysis, outperformed nine baseline algorithms in terms of average Recall, Precision, and \F~scores.
    A promising extension of our work involves integrating more information to enhance prediction reliability. Another extension of the work is to perform our proposed itinerary prediction algorithm on a large set of datasets.
    We will also perform ablation experiments to verify the experimental results.

\section*{Acknowledgment}
  \noindent{\small This research is funded in part by the Singapore University of Technology and Design under grant RS-MEFAI-00005-R0201. 
  The computational work was partially performed on resources of the National Super-Computing Centre.
      The computational work was partially performed on resources of the \href{https://www.socialai.studio/}{Social AI Studio}, \href{https://dai.sutd.edu.sg/}{DAI}.}
  The computational work was partially performed on resources of the National Supercomputing Centre, Singapore



\bibliographystyle{unsrt}  
\bibliography{arxiv}

\end{document}